%% file: ms.tex
\documentclass[10pt,conference,a4paper]{IEEEtran}
\usepackage{url}
\usepackage{acronym}
\usepackage{multirow}
\usepackage{adjustbox}
\usepackage{graphicx}
\usepackage{amsmath}
\usepackage{subcaption}
\usepackage{filecontents}
\usepackage{colortbl}
\usepackage{textcomp} 
\usepackage{enumitem}
\usepackage[ruled]{algorithm2e}
\usepackage{threeparttable}
\usepackage{tablefootnote}
\usepackage{lipsum}
\usepackage{todonotes}
\usepackage[mathlines,switch]{lineno}
\usepackage{pifont}
\usepackage[english]{babel}
\usepackage{tikz}
\usetikzlibrary{positioning}
\usepackage[backend=biber, 
            style=numeric-comp,
            natbib=true,
            doi=false,
            isbn=false,
            url=true,
            maxnames=3, 
            maxcitenames=3,
            giveninits=true,
            bibencoding=utf8,
            sorting=none]{biblatex}
\begin{document}

\title{A Subjective Dataset for Multi-Screen Video Streaming Applications}
\author{
    \IEEEauthorblockN{ Nabajeet Barman\IEEEauthorrefmark{1}, Yuriy Reznik\IEEEauthorrefmark{2}, Maria G. Martini\IEEEauthorrefmark{3}}
        \IEEEauthorblockA{\IEEEauthorrefmark{1}Brightcove UK Ltd, London, United Kingdom, n.barman@ieee.org}
    \IEEEauthorblockA{\IEEEauthorrefmark{2}Brightcove Inc, Seattle, USA, yreznik@brightcove.com}
        \IEEEauthorblockA{\IEEEauthorrefmark{3}Kingston University, London, United Kingdom, m.martini@kingston.ac.uk}
        }


\maketitle

\begin{abstract}
In modern-era video streaming systems, videos are streamed and displayed on a wide range of devices. Such devices vary from large-screen UHD and HDTVs to medium-screen Desktop PCs and Laptops to smaller-screen devices such as mobile phones and tablets. It is well known that a video is perceived differently when displayed on different devices. The viewing experience for a particular video on smaller screen devices such as smartphones and tablets, which have high pixel density, will be different with respect to the case where the same video is played on a large screen device such as a TV or PC monitor. Being able to model such relative differences in perception effectively can help in the design of better quality metrics and in the design of more efficient and optimized encoding profiles, leading to lower storage, encoding, and transmission costs.

However, to the best of our knowledge, open-source datasets providing subjective scores for the same content when viewed on multiple devices with different screen sizes do not exist, thus limiting a proper evaluation of the existing quality metrics for such multi-screen video streaming applications. This paper addresses this research gap by presenting a new, open-source dataset consisting of subjective ratings for various encoded video sequences of different resolutions and bitrates (quality) when viewed on three devices of varying screen sizes: TV, Tablet, and Mobile. Along with the subjective scores, an evaluation of some of the most famous and commonly used open-source objective quality metrics is also presented. It is observed that the performance of the metrics varies a lot across different device types, with the recently standardized ITU-T P.1204.3 Model, on average, outperforming their full-reference counterparts. The dataset consisting of the videos, along with their subjective and objective scores, is available freely on Github\footnote{https://github.com/NabajeetBarman/Multiscreen-Dataset}.
\end{abstract}



\IEEEpeerreviewmaketitle

\section{Introduction}

The past two decades have seen a tremendous improvement in streaming technologies, resulting in the growth and popularity of streaming services delivering media over the internet, often referred to as Over-the-top (OTT) applications. One of the major reasons behind the popularity of such services is the evolution of the concept of adaptive streaming, where media bitrate is adapted based on the (fluctuating) network bandwidth, thus allowing the users to experience a smooth playback on their device. The concept itself is not new, with the first commercial product built on ABR principles being the RealNetworks system G2 released in July 1998~\cite{Reznik_SMPTE_MassScale}. Nowadays, HTTP-based Adaptive Streaming (HAS) protocols such as HLS~\cite{HLS} and DASH~\cite{DASH} have become the de-facto streaming technology. 

The advancement in encoding, transmission, and playback capabilities have resulted in the growing popularity of such streaming services, changing how people consume content. For example, previously, in traditional broadcasting, the video was usually delivered to a single device, a TV, with most of them of similar screen size. However, in today's modern era of streaming, videos are delivered to many different devices with different playback capabilities~\cite{Reznik2021_PercepWebStreaming}. The distribution of playback devices can range from large screens, e.g., 75$''$ or 65$''$ TVs, to medium and small-sized devices such as Tablets and Mobile phones. Not limited to on-demand video streaming, there is also an increasing usage of video conferencing solutions such as Google Meet, Zoom, and Microsoft  Teams. Such online meetings can now take place over multiple screens, such as on a large TV in a conference room or on a laptop or mobile phone, depending on the attendees' preference. In some cases, the users even switch devices during a session, with an expectation of the same consistent quality of experience. 

Such a complex ecosystem of videos being streamed to multiple devices with different screen sizes and form factors (pixel density and viewing distance) brings many inherent challenges, from a selection of codecs and profiles to encoding bitrates and resolutions, to make sure the video playback is supported on multiple devices. This often necessitates the generation of multiple copies of the same content, leading to increased storage and transcoding costs, a problem faced by almost all online video providers. The availability of generalized parametric models, which can help model the relative difference in perception due to the use of multiple screens, can help in a better understanding of the performance of the streaming systems, which in turn, can be used to design better, more optimal content and context-aware streaming applications, thus reducing the number of generated streams per content and hence, the encoding and streaming costs~\cite{Reznik_EUVIP22}.

\subsection{Prior Work}

The effect of display-related parameters such as rendered image/video size, image/video resolution, viewing distance, and display pixel density, among others, is well known and studied in the literature~\cite{Jesty1958,wr1989_model,Lund1993,Barten}. For example, Catellier \textit{et al} in~\cite{catellier2012_mobile} demonstrated that lower quality content is usually rated slightly higher on displays with high pixels per degree. Similarly, in an ongoing standardization effort ITU-T Rec. SG12-TD1612\cite{ITU_display}, it is shown that only 40$''$ TV monitors showed noticeable quality improvement when video resolution was 1080p as compared to 720p. No noticeable improvement, however, was observed for smaller screen devices such as mobile and tablets. In the same study, no noticeable improvement in perceptual quality was observed between FHD and UHD video sequences, even when considering a 75$''$ TV screen. Authors in \cite{sugito8kVVC} performed subjective evaluation experiments on 8K videos encoded using VVC~\cite{h266} under seven viewing conditions (a combination of various screen sizes and viewing distances). They also found that the perceived quality is higher in smaller screens or at increased viewing distances. However, the subjective test results are not made available to the public. 

However, most QoE models and metrics currently either ignore the effects introduced due to the differences in devices (e.g, PSNR and SSIM~\cite{SSIM}) or require recomputation for each device/display type (e.g, VMAF~\cite{NetflixVMAF_Github} and ITU-T P.1204.3~\cite{rao2020p1204}). In an effort towards the design of a content-independent model of variation of subjective scores across different devices, viewing modes, and viewing distances, authors in Ref.~\cite{balu2020_displays} proposed a content-independent model of variation of subjective scores across different devices, viewing modes, and viewing distances. The authors use a power-law function to derive a relationship between the subjective scores (MOS and DMOS) and display size and viewing distances using the ratio of the device pixels-per degree (PPD). Similarly, in \cite{Zou2016ImageQuality}, the authors model the perceived image quality as a function of coded image quality using a power law whose parameters are dependent on the downsampling factor while incorporating the device characteristics in terms of effective-displayed pixels-per-inch (ED-PPI). More recently, Reznik \textit{et al.} in ~\cite{Reznik_EUVIP22} presented simple parametric models for predicting visual quality scores on different devices in multi-screen systems considering the viewing setup, resolution of the projected video, size of the display, and viewing distance. Using different datasets, the authors demonstrated the utility of the proposed models to effectively capture the relative perceptual differences introduced by using different devices. 

\subsection{Existing Challenges}

As discussed in the previous section, while there have been few works towards the design of QoE models capturing the effects due to differences in perception due to different devices (screen), the performance evaluation of such models is often performed using a combination of different datasets, each considering a single device type. Such datasets, due to the use of different content as well as test subjects, environment, rating scale, etc., fail to capture the relative differences in perceived quality between different devices. Also, none of these works includes an open-source subjective multi-screen dataset containing MOS capturing the relative perceptual differences of the same content when viewed on multiple devices/screens. This paper presents a dataset addressing this research gap. 

\subsection{Contributions}

We present in this paper an open-source dataset consisting of subjective scores of various videos displayed on three different devices, TV, Tablet, and Mobile. The source video sequences were encoded in three different resolution-bitrate pairs, which were then played and rated by test subjects on three devices of different screen sizes. The source and encoded video sequences, along with mean opinion score ratings, are provided as part of the dataset. In addition, we also present a discussion on the performance of four widely used Full-Reference and the recently standardized ITU-T Rec. P1204.3 bitstream-based No-Reference model~\cite{rao2020p1204} on this dataset. We believe the presented dataset will be of much use to the research community to design better metrics and models for modern-era streaming applications delivering videos to multiple devices.

A few possible research areas where this dataset can contribute are as follows: 
\begin{itemize}
\item Gaining insight into the effect of different screen sizes on the viewing experience of the end user;
\item Evaluating the performance of existing VQA algorithms and methods (e.g., quality metrics) considering a multi-screen streaming environment;
\item Design of new parametric models and metrics taking into account differences in perception due to varied screen sizes;
\item Cross-lab evaluation and comparison of subjective quality assessment studies.
\end{itemize}
\begin{figure}[t!]
\begin{center}
    \includegraphics[width=0.70\linewidth]{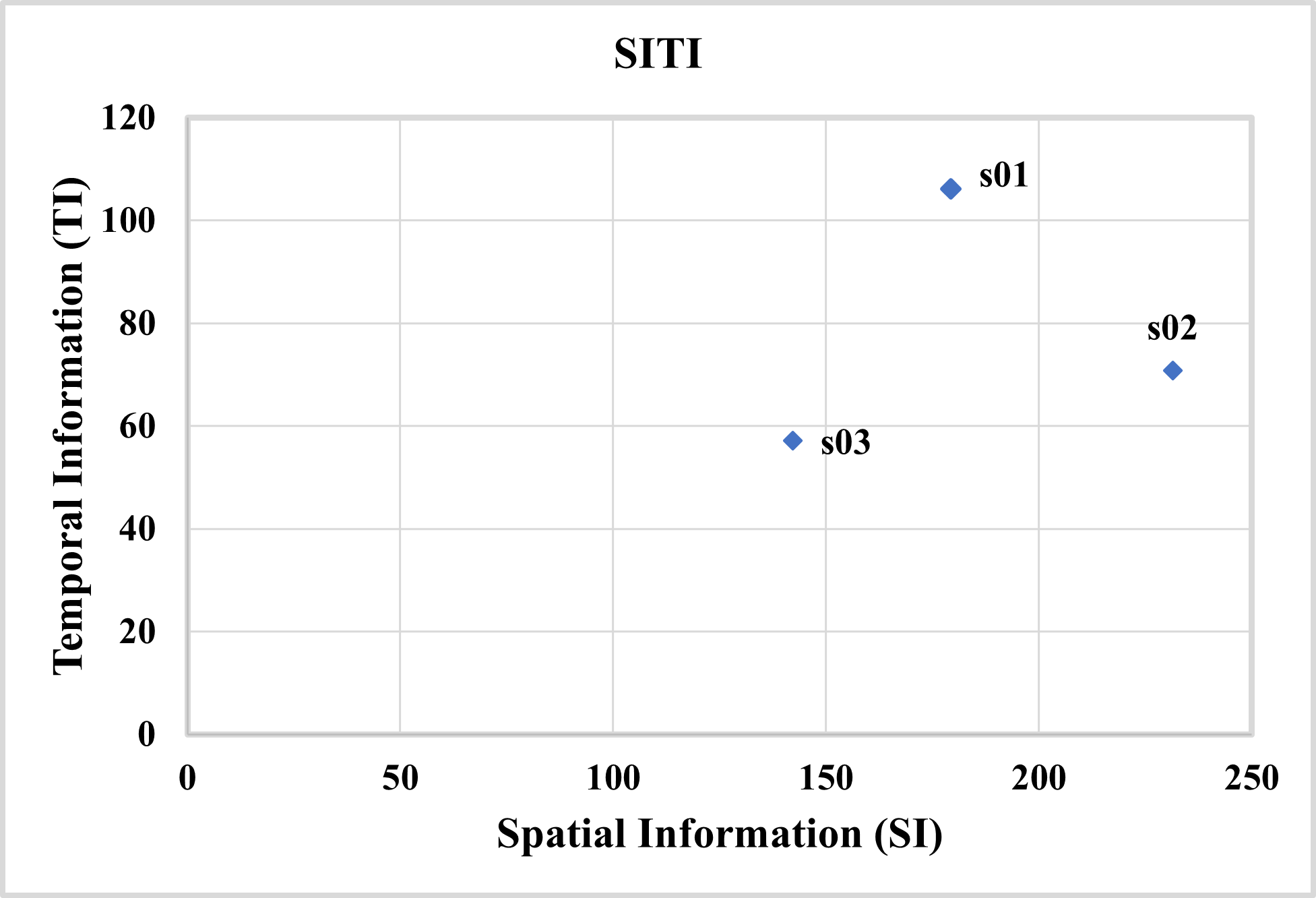}
\end{center}
\caption{ 
SI vs TI plot for the three source sequences.}
\label{fig:siti}
\end{figure}
The rest of the paper is organized as follows. 
Section~\ref{sec:testSeq} presents the overview of the source sequences, their complexity, and encoding settings used to obtain the test sequences. Section~\ref{sec:subj_exp} discusses the subjective test procedure. Section~\ref{sec:res} presents the subjective and objective evaluation results. Section~\ref{sec:concandfw} concludes the paper.

\section{Test Sequences and Encoding Settings} \label{sec:testSeq}

\begin{figure}[t!]
\begin{center}
    \includegraphics[width=1\linewidth]{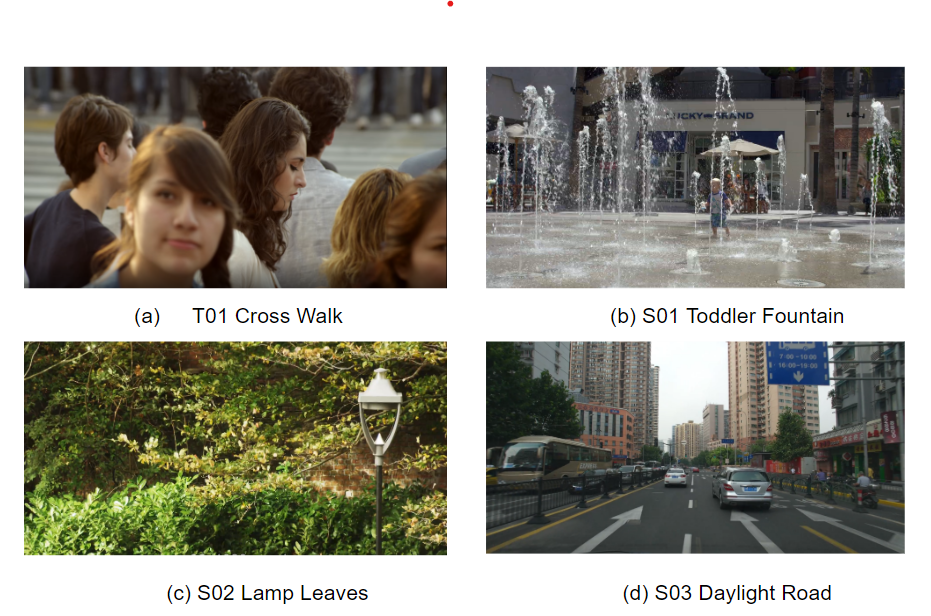}
\end{center}
\caption{Screenshots of the source videos used in this dataset. Sequence (a) is used for training, while the rest three sequences, (b)-(d), are used in the actual test.}
\label{fig:screenshots}
\end{figure}

\subsection{Source Sequences}

We selected four source sequences from the BVI-CC dataset~\cite{BVI_CC}, which is a well-known and widely used dataset for research on video compression and quality assessment. Figure~\ref{fig:screenshots} above shows the screenshots of the videos used in this work. Sequence (a) was used during the training session, while sequences (b), (c), and (d) were used in the actual subjective test. To validate that the selected videos are of varied complexity and representative of what is observed in real-world applications, we calculated the Spatial Information (SI) and Temporal Information (TI) values of the three source sequences. SI and TI values can be used as an approximate measure of complexity, with high SI and high TI values representing a high level of complexity~\cite{barman2019SITI}. Figure~\ref{fig:siti} shows the Spatial Information (SI) vs Temporal Information (TI) plot

for all three video sequences. Based on the figure, it can be observed that the selected source sequences are of hard to medium complexity and should be representative of the challenging contents of any major streaming application. 

\subsection{Encoding Settings}

The encoding of the source files was performed by a commercial HEVC encoder, Zencoder~\cite{zencoder} using the Context Aware Encoding (CAE) setting~\cite{cae}. CAE analyzes each source video and intelligently builds a custom bitrate ladder (set of renditions) for each piece of content while also taking into account constraints associated with the delivery network and device being used to view the content. However, to make sure that the generated profiles are consistent across the three different sources, we introduce a set of constraints on framerate (= 60, same as source), resolutions (540p, 720p, and 1080p), and quality (roughly high, medium and low). The three video qualities (high, medium, and low) are generated using the CAE's “quality” setting, which adjusts the quality-bitrate trade-off to obtain the encoded video representation. The exact "quality" setting value was obtained based on a visual inspection of the obtained encoded sequences. More details about the encoded sequences, along with the encoded video sequences' bitrates, are available in the dataset. More details about the encoded sequences, along with the encoded video sequences' bitrates, are available in the dataset.

\begin{table}[t!]
\centering
\caption{Summary of video encoding settings.}
\label{table:EncodingSummary}
\renewcommand{\arraystretch}{1.2}
\begin{tabular}{|l|l|}
\hline
\textbf{Parameter}                   & \textbf{Value}           \\ \hline
Duration                             & 5 sec                    \\ \hline
Resolution                           & 1080p, 720p, 540p        \\ \hline
Dynamic Range                        & SDR (BT.709)        \\ \hline
Framerate                            & 60                       \\ \hline
Encoder                              & Zencoder~\cite{zencoder}                \\ \hline
Encoding Mode                        & Brightcove's CAE~\cite{cae}        \\ \hline
Codec/Profile                        & HEVC (Profile: Main)     \\ \hline
\end{tabular}
\end{table}

\section{Subjective Experiments} \label{sec:subj_exp}

In the absence of any existing recommendations on subjective assessment of content in a multiple device setting, we relied on existing relevant ITU-T Standards (ITU-T BT500-14~\cite{bt500-14}, and ITU-T P.913~\cite{p913}) to design the subjective test setup for conducting this study. We next discuss the selected devices, test environment, and the test methodology, along with details about the test participants.  

\subsection{Test Devices}

After careful consideration based on the availability, screen size, and display type, the following three devices were shortlisted and used in this study:
\begin{enumerate}
    \item \textit{TV}: 55$''$ Samsung Q9F Flagship QLED 4K Certified Ultra HD Premium HDR 2000 Smart TV powered by HDR10+			
    \item \textit{Tablet}: 12.4-inch Samsung Tablet with 2800 x 1752 (WQXGA+) resolution super amoled display			
    \item \textit{Mobile}: 6.1$''$ Samsung Galaxy S21 5G smartphone with device resolution of 2400x1080 pixels.
\end{enumerate}

The display for all three devices was set to the default manufacturer settings, and HDR and other related enhancements, if any, were switched off. More details about the devices and their settings, along with a link to their technical specifications, are available as part of the dataset.

\subsection{Test Environment}

\begin{figure}[t!]
\begin{center}
    \includegraphics[width=1\linewidth]{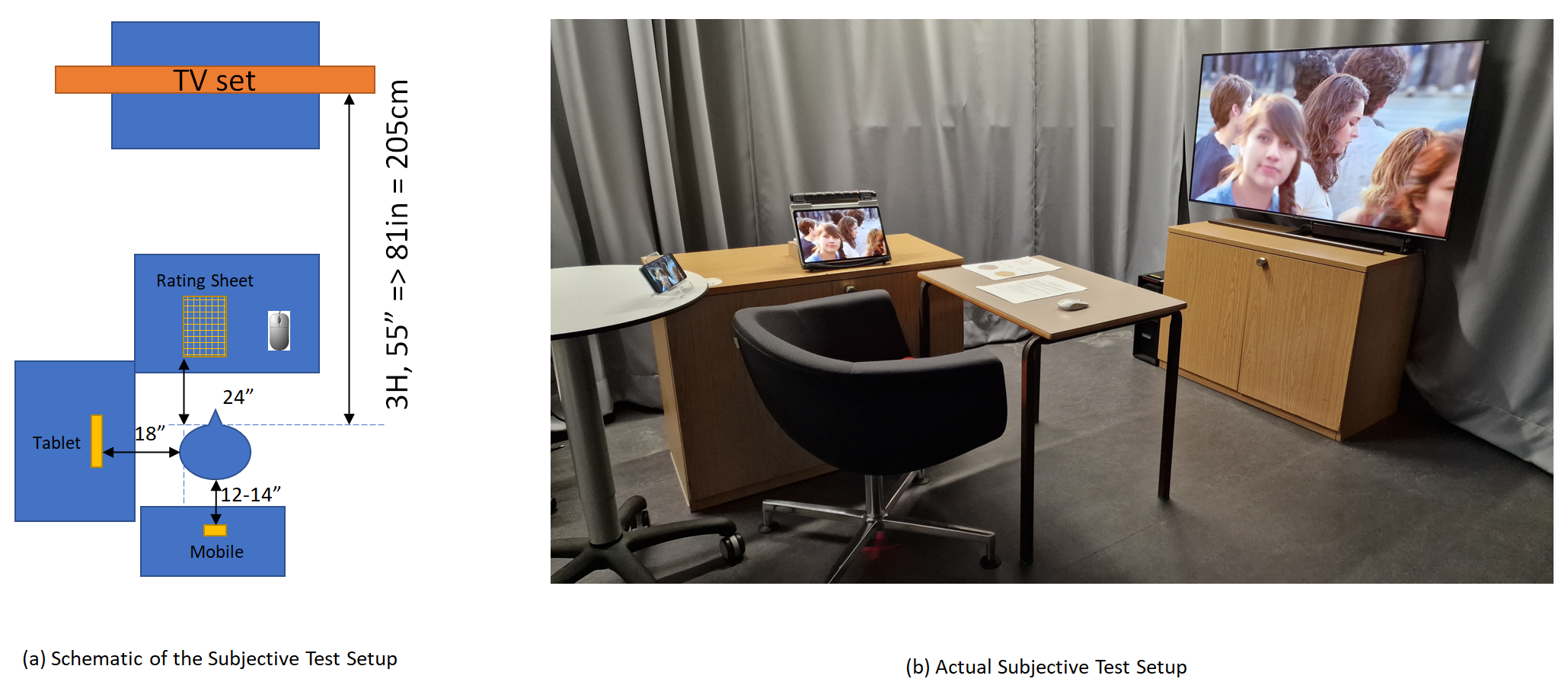}
\end{center}
\caption{Subjective test lab setup: (a) Schematic (b) Actual.}
\label{fig:test_setup}
\end{figure}

The subjective test was conducted in the Centre for Augmented and Virtual Environments (CAVE) at Kingston University, London, United Kingdom, which is specifically designed for conducting subjective tests adhering to ITU-R Rec. BT.500 standard \cite{bt500-14}. Figure~\ref{fig:test_setup} illustrates both the schematic and the actual subjective lab setup used to conduct this study. The schematic illustrates the viewing distance of the devices/screens from the observer: TV= 3H (=81$''$), Tablet= 18$''$, Mobile= 12-14$''$. The values are selected based on well 
known statistics
as reported in other relevant literature~\cite{Reznik_EUVIP22,balu2020_displays,p913,bt500-14,Amirpour2021ViewingDistance,SPIE2015ViewingDistance,MobileVQDB}. The room lighting was maintained between 60
and 110 lux
and it was made sure that there was no glare on any screen due to the room lights or light from other screens. The test environment was quiet, without any distractions or external noise. No other person was in the room during the subjective test (except for the test supervisor, as required by Kingston University regulations). 

\subsection{Experimental Procedure}

The testing was performed according to ITU-R Rec BT.500-14 methodology~\cite{bt500-14}. The 5-point Absolute Category Rating ACR scale was used for the quality assessment.
The test participants first filled in a consent form followed by a pre-test questionnaire answering some questions about demographics (age, gender, and viewing habits). This was followed by a training session where they were introduced to three different quality 
levels 
(low, medium, and high)
via the 
training video sequence T01 Cross Walk (see Figure~\ref{fig:screenshots}) on the three devices. This allowed them to get used to the test procedure, interface, and video playback. The subjects were asked to evaluate the “picture” quality and ignore any particular likeness for the video, as is done in most QoE studies. The subjects were also 
informed of
the motivation and objective of this study, followed by a discussion of any queries they had related to the procedure or the test in general. 

Before starting the subjective test, the test subjects were first seated and then the device placement/table for smartphone and tablet was adjusted for each test participant to ensure they were at the required distance (from the center of the screen of the devices to the participants' eye level) as shown in Figure~\ref{fig:test_setup}. They were instructed to maintain the required viewing distance during the course of the subjective test. The test was started by the test participants when they felt comfortable. They were allowed to take breaks whenever they wanted, stop the study if they felt discomfort, or ask questions to the supervisor. At the end of the test, the lighting was adjusted, and the participants filled in the post-test questionnaire by answering four short questions describing their experience. 

\subsection{Video Playback}

For video playback, the latest version of the VLC player
was used. The video to the TV was delivered from a Windows PC connected over HDMI with VLC Version 3.0.18 for Windows installed. VLC 3.5.3 for Android was installed on the Tablet and Mobile via Google Playstore. The VLC player settings were changed to ensure that the video is always played in full screen and no additional video-related information was displayed during the video playback. The upscaling of videos to the display resolution was performed by the VLC player. The settings were kept the same across all three devices and the test subjects were asked to watch the same video on all three devices in order of their preference. 
However, as the sequences were only five seconds long, the subjects were allowed to repeat a particular sequence as many times as they wanted before they agreed on the rating. Once happy with their judgment of quality, they were asked to rate the video sequence on a scale from 1-5 for each device separately.

\subsection{Test Participants Demographics}

A total of 26 test participants took part in the study, with 17 male and nine female test participants. The participants mostly included current and past Kingston University students, along with a few academics and other university staff. This allowed us to capture a wide range of opinions and not restrict them to a particular demographic, as done in many other subjective studies (e.g., including only young students). They were almost equally distributed 
among three age groups: 18-24, 25-34, and 35-44, with nine of them 
having participated
in a subjective test before. Before starting the test, the participants were tested for visual acuity and color blindness using Snellen charts and Ishihara plates, respectively. All had (corrected) normal vision, and no color blindness was observed among the test participants. Additional details about the subjective experiments such as individual subjective ratings for all three devices, statistical analysis on the user ratings, and additional demographic information about the test participants including viewing habits, etc. are not presented here due to page limitation but are available in the GitHub repository.

\begin{figure}[t!]
    \centering
    \begin{subfigure}[t]{0.85\linewidth}
    \includegraphics[width=1\linewidth]{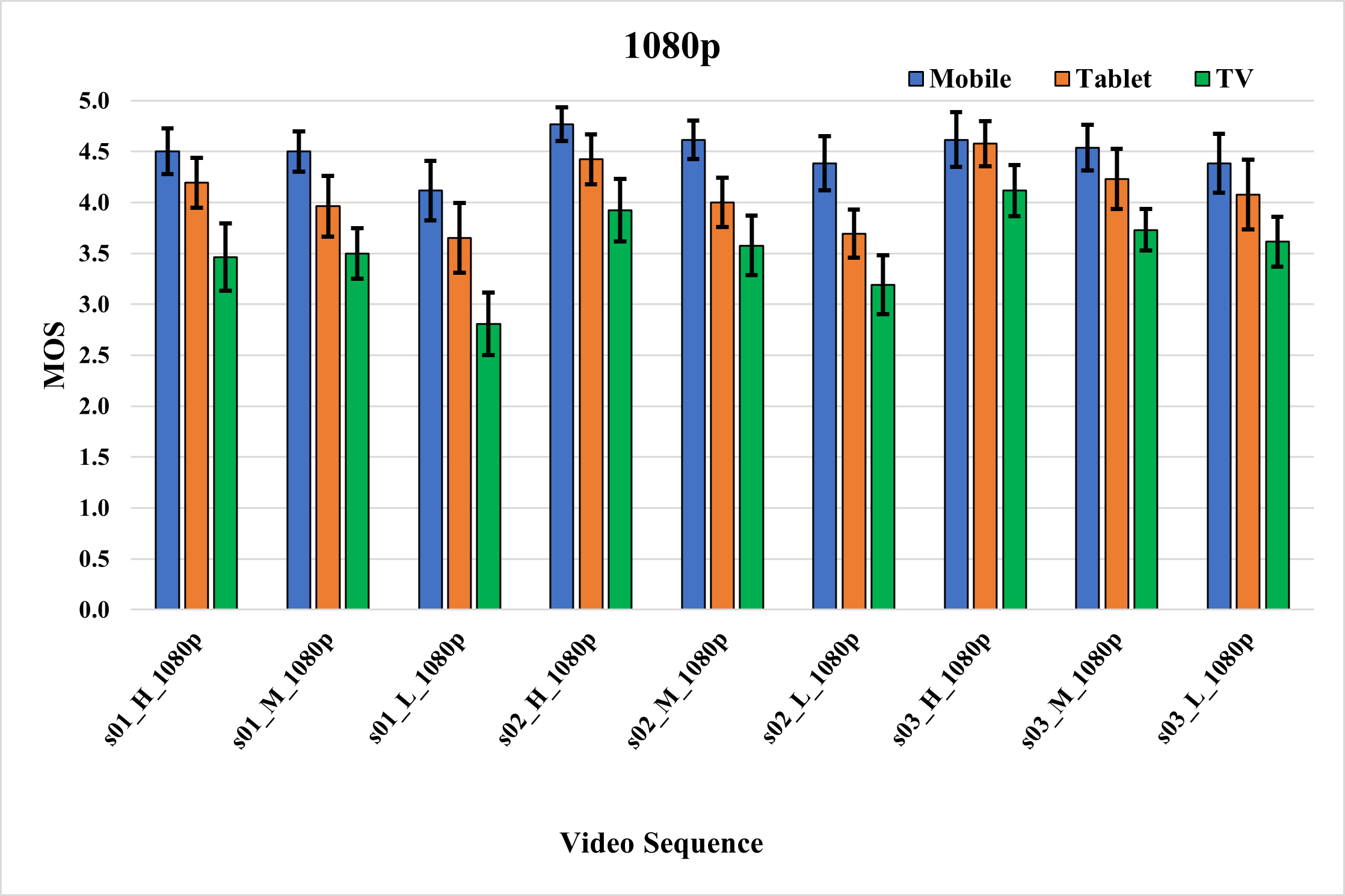}
    \end{subfigure}

    \begin{subfigure}[t]{0.85\linewidth}
    \centering
    \includegraphics[width=1\linewidth]{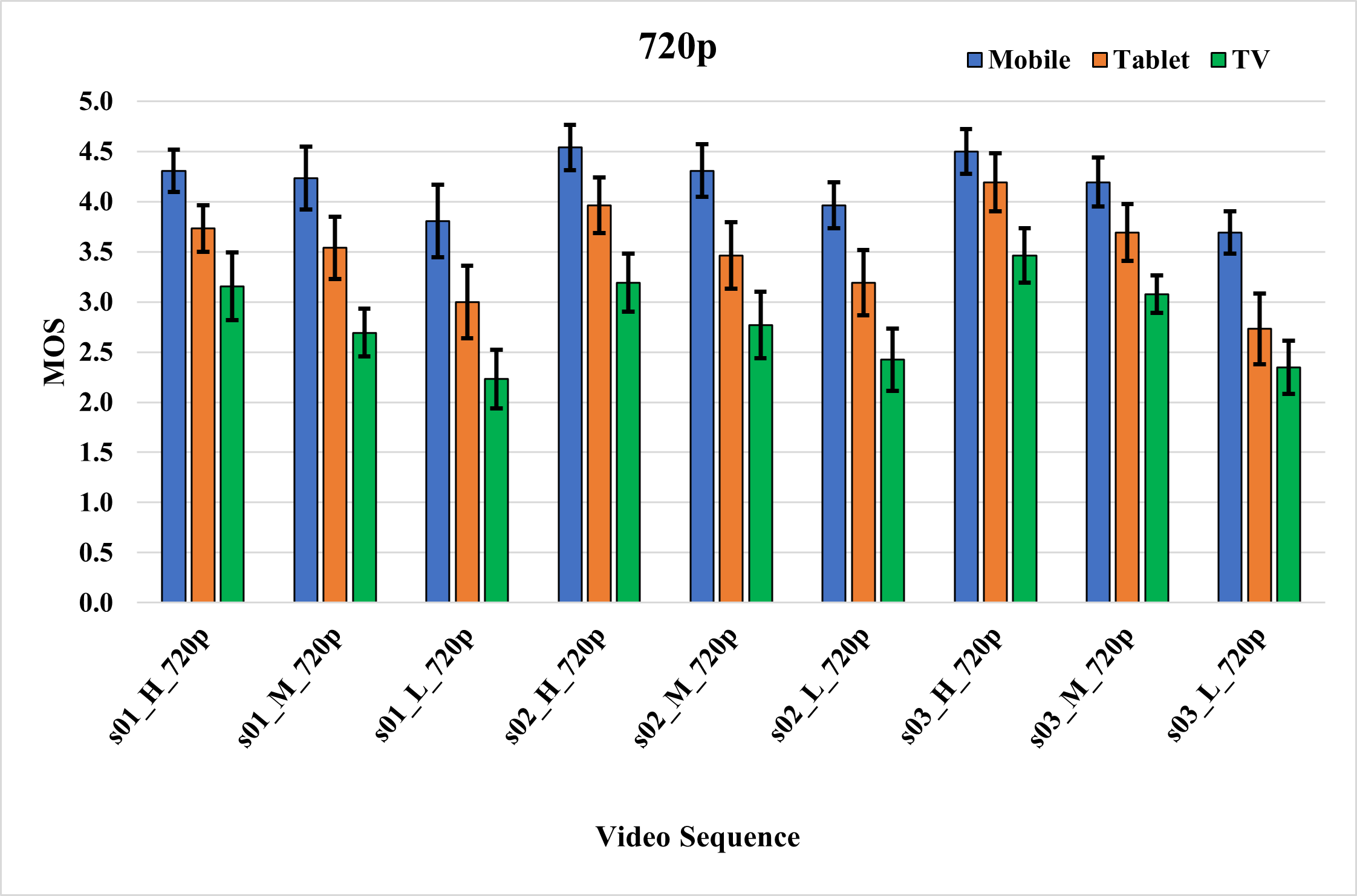}
    \end{subfigure}
    
    \begin{subfigure}[t]{0.85\linewidth}
    \centering
    \includegraphics[width=1\linewidth]{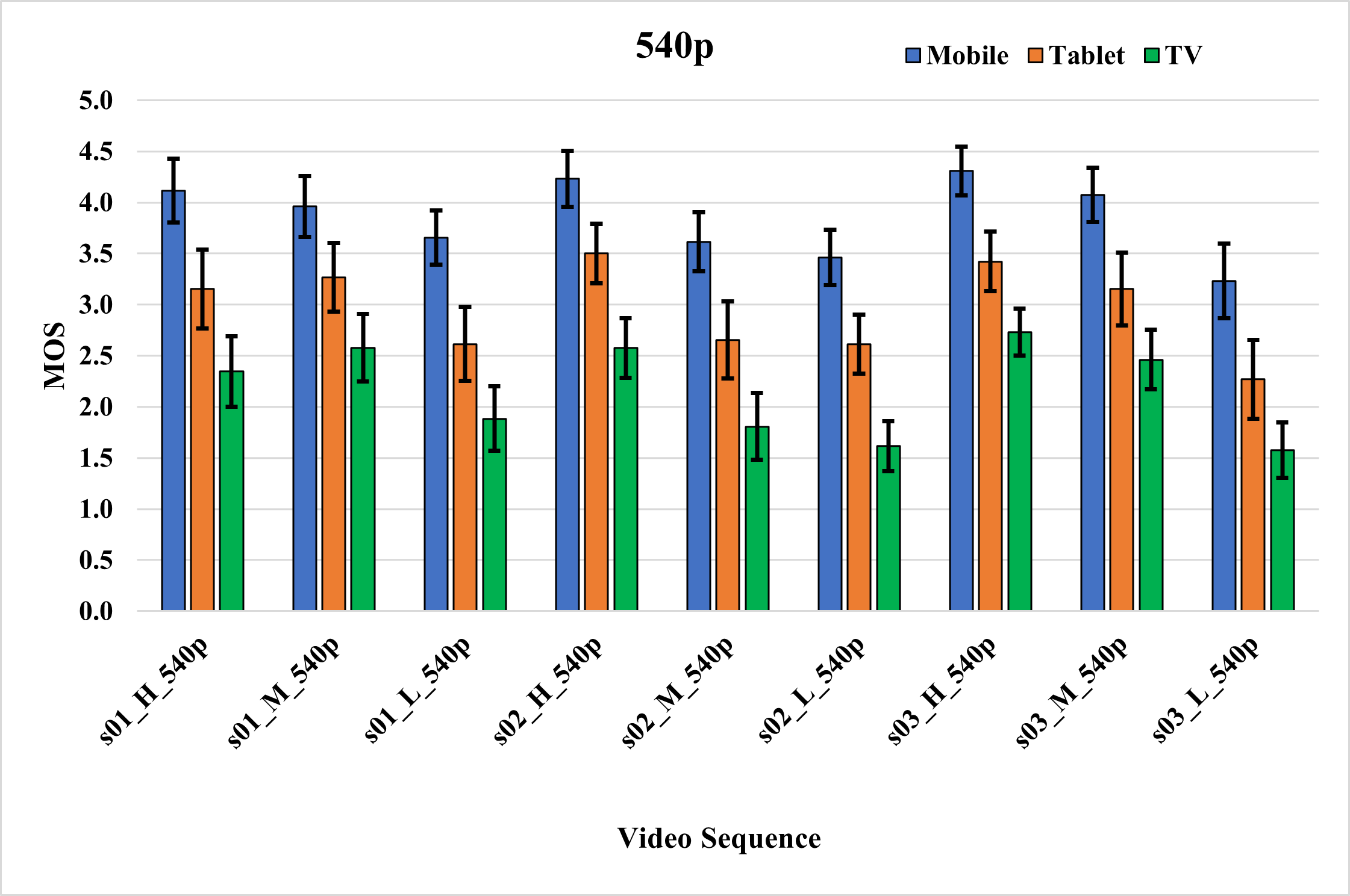}
    \end{subfigure}
    \captionsetup{font=small}
     \caption{MOS (with 95\% confidence interval) for all video sequences for all three different devices considering three different resolutions.}
    \label{fig:mosPlots}
\end{figure}

\section{Results and Discussions} \label{sec:res}

\subsection{Subjective Assessment}

Figure~\ref{fig:mosPlots} presents the bar plots showing the average of the opinion scores, also commonly known as Mean Opinion Scores (MOS), along with the 95\% confidence interval. The plots are shown separately for each encoded resolution for all three devices. Based on the plots, the following can be observed:
\begin{enumerate}
    \item A particular video sequence is almost always rated the highest on a mobile device. Tablet devices receive a slightly lower rating, while on a larger screen, such as a TV, the rating for the same video can be quite low.
    \item In general, even the lowest quality videos were rated at least ``fair" (3 MOS) on a mobile device.
    \item The difference in quality for the 1080p resolution renditions, considering each device individually, 
    is more prominent for lower-quality encodes as compared to medium and high-quality encodes of the same video. 
    \item At lower encoded resolutions (720p and 540p), the difference in quality when considering different devices is more prominent for 540p resolution, especially for mobile and tablet devices.
\end{enumerate}

\subsection{Objective Assessment}

Since subjective tests are not practical for evaluating real-world applications, objective quality assessment metrics are of high interest to both academia and industry~\cite{barman2018PV}. Towards this end, we evaluated the performance of four of the most commonly used full reference (FR) VQA metrics on the dataset: PSNR, SSIM~\cite{SSIM}, VIF~\cite{vif} and VMAF~\cite{NetflixVMAF_Github}. For VMAF, only the default model (1080p TV) was considered for performance evaluation across all three devices as the VMAF phone model could not be evaluated due to an issue with the \textit{libvmaf} implementation in FFmpeg. However, the dataset repository will be updated later to include the VMAF phone model scores. In addition, we also evaluated the ITU-T P.1204.3 model~\cite{rao2020p1204}, which is a no-reference bitstream-based model. The model supports two variants: mobile/tablet and TV models, both of which were computed separately for performance evaluation for the corresponding device type. 

Table~\ref{tab:corr} shows the Pearson and Spearman correlation coefficients along with RMSE scores for the five quality metrics. The values are obtained after performing a first-order linear mapping of the scores to account for the differences in scale between the subjective and objective quality ratings. As expected, PSNR and SSIM do not have a high correlation with subjective scores obtained for the three devices. Interestingly, VIF performs quite close to VMAF in almost all three cases with it performing the best for the Mobile and TV devices in terms of SROCC scores. Among all the five metrics considered here, it can be observed that, on average, across all devices, P.1204.3 performs the best, considering all three performance measures. It is quite remarkable since, compared to the other four metrics, which are pixel-based full-reference models, P.1204.3 is a no-reference bitstream-based model making it suitable for real-world QoE monitoring. 

We discuss next the performance of a particular metric across different devices considering the various performance measures. In general, when considering different device types, the performance of simple metrics such as PSNR and SSIM improves as we move from smaller-screen devices (Mobile) to larger-screen devices (TV). However, interestingly enough, for more complex metrics such as VIF, VMAF and P.1204.3, in terms of PLCC and SROCC scores, we observe an improved performance for Tablet device cases as compared to Mobile and TV devices. However, in terms of RMSE scores, we see consistent behaviour across all classes of metrics with an improvement in performance as we move from smaller screen devices to larger screen devices. The difference in the behaviour of a metric across different performance measures for different devices indicates the need for consideration of various performance measures for a fair and complete comparison of different quality metrics. 

\input{Tables/corr.tex}

\section{Conclusions and Future Work} \label{sec:concandfw}

We presented in this paper the first-ever open-source dataset capturing the relative differences in perception when watching the same video on three different devices, mobile, tablet, and TV. The dataset consists of MOS scores separately for all three devices and hence can also be used in conjunction with other existing subjective datasets to study other aspects such as encoding performance, codec comparison, etc. We also presented the performance of the four most commonly used FR metrics and the recently standardized ITU-T Rec P.1204.3, a no-reference bitstream-based model. Our results in terms of PLCC, SROCC, and RMSE scores indicate a very good performance of the ITU-T P.1204.3 standard, on par with full-reference metrics across different performance measures. 

The consistent performance of P.1204.3 across device types indicates that usage of contextual parameters such as device type has the potential of enhancing the prediction accuracy of quality models. However, the need to compute metrics such as VMAF and P.1204.3 for each device type separately adds additional complexity. 
A better approach would be to design parametric models which can be adapted to take into account the differences in viewing setup (device type, viewing distance, etc) without the need for re-computation for each device type~\cite{Reznik_EUVIP22}.
In the future, we plan to extend the study to include more source sequences of higher resolution (8K), bit-depth (10-bit), dynamic range (HDR) and other video compression standards.


\section*{Acknowledgements}

Nabajeet Barman would like to thank all the test participants, Dr Arslan Usman, and the KU SEC-IT team.

\section*{References}
{
\begingroup
    \renewcommand{\bibfont}{\small}
    \printbibliography[heading=none]
\endgroup
}

\end{document}

%% file: Tables/corr.tex
\begin{table*}[t!]
\centering
\caption{Comparison of the performance of various VQA metrics in terms of PLCC, SROCC, and RMSE scores. The best-performing metric for each device is shown in bold.}
\label{tab:corr}
\renewcommand{\arraystretch}{1.5}
\resizebox{\textwidth}{!}{%
\begin{tabular}{|l|ccccc|ccccc|ccccc|}
\hline
\multicolumn{1}{|c|}{\multirow{2}{*}{\textbf{Device}}} & \multicolumn{5}{c|}{\textbf{PLCC}}                                                                                                                                  & \multicolumn{5}{c|}{\textbf{SROCC}}                                                                                                                                  & \multicolumn{5}{c|}{\textbf{RMSE}}                                                                                                                                  \\ \cline{2-16} 
\multicolumn{1}{|c|}{}                                 & \multicolumn{1}{c|}{\textbf{PSNR}} & \multicolumn{1}{c|}{\textbf{SSIM}} & \multicolumn{1}{c|}{\textbf{VIF}} & \multicolumn{1}{c|}{\textbf{VMAF}} & \textbf{P1204.3} & \multicolumn{1}{c|}{\textbf{PSNR}} & \multicolumn{1}{c|}{\textbf{SSIM}} & \multicolumn{1}{c|}{\textbf{VIF}}  & \multicolumn{1}{c|}{\textbf{VMAF}} & \textbf{P1204.3} & \multicolumn{1}{c|}{\textbf{PSNR}} & \multicolumn{1}{c|}{\textbf{SSIM}} & \multicolumn{1}{c|}{\textbf{VIF}} & \multicolumn{1}{c|}{\textbf{VMAF}} & \textbf{P1204.3} \\ \hline
\textbf{Mobile}                                        & \multicolumn{1}{c|}{0.56}          & \multicolumn{1}{c|}{0.68}          & \multicolumn{1}{c|}{0.81}         & \multicolumn{1}{c|}{0.83}          & \textbf{0.86}    & \multicolumn{1}{c|}{0.56}          & \multicolumn{1}{c|}{0.71}          & \multicolumn{1}{c|}{\textbf{0.82}} & \multicolumn{1}{c|}{0.80}          & \textbf{0.82}    & \multicolumn{1}{c|}{0.32}          & \multicolumn{1}{c|}{0.29}          & \multicolumn{1}{c|}{0.23}         & \multicolumn{1}{c|}{0.22}          & \textbf{0.20}    \\ \hline
\textbf{Tablet}                                        & \multicolumn{1}{c|}{0.65}          & \multicolumn{1}{c|}{0.72}          & \multicolumn{1}{c|}{0.85}         & \multicolumn{1}{c|}{0.86}          & \textbf{0.89}    & \multicolumn{1}{c|}{0.65}          & \multicolumn{1}{c|}{0.76}          & \multicolumn{1}{c|}{0.85}          & \multicolumn{1}{c|}{0.86}          & \textbf{0.89}    & \multicolumn{1}{c|}{0.49}          & \multicolumn{1}{c|}{0.44}          & \multicolumn{1}{c|}{0.34}         & \multicolumn{1}{c|}{0.32}          & \textbf{0.29}    \\ \hline
\textbf{TV}                                            & \multicolumn{1}{c|}{0.68}          & \multicolumn{1}{c|}{0.77}          & \multicolumn{1}{c|}{0.85}         & \multicolumn{1}{c|}{\textbf{0.86}} & 0.85             & \multicolumn{1}{c|}{0.65}          & \multicolumn{1}{c|}{0.77}          & \multicolumn{1}{c|}{\textbf{0.85}} & \multicolumn{1}{c|}{0.82}          & 0.81             & \multicolumn{1}{c|}{0.54}          & \multicolumn{1}{c|}{0.47}          & \multicolumn{1}{c|}{0.39}         & \multicolumn{1}{c|}{\textbf{0.38}} & 0.39             \\ \hline
\end{tabular}}
\end{table*}